\PassOptionsToPackage{table}{xcolor}
\documentclass[sigconf]{acmart}
\renewcommand\footnotetextcopyrightpermission[1]{}
\usepackage[colorinlistoftodos]{todonotes}

\usepackage{xcolor}

\usepackage{algorithm}
\usepackage{amsfonts}
\usepackage{amsmath}
\usepackage{amsthm}
\usepackage{array}
\usepackage[T1]{fontenc}
\usepackage{bbm}
\usepackage{booktabs}
\usepackage{caption}
\usepackage{color}
\usepackage{csquotes}
\usepackage{enumitem}
\usepackage{epigraph} 
\usepackage{fontawesome5}
\usepackage{graphicx}
\graphicspath{{figures/}}
\usepackage{hhline}
\usepackage{hyperref}
\usepackage[capitalise,noabbrev]{cleveref}
\usepackage{ltablex}
\keepXColumns
\usepackage{makecell}
\usepackage{mathtools}
\usepackage{mdframed}
\usepackage{multirow}
\usepackage{pifont}
\usepackage{tikz}
\usepackage{soul}
\usepackage{stmaryrd}
\usepackage{subcaption}
\usepackage{url}
\usepackage{wrapfig}

\newif\ifcomments
\commentstrue
\newcommand{\kl}[1]{\ifcomments\textit{\color{magenta}[KL] : #1}\fi}
\newcommand{\djw}[1]{\ifcomments\textit{\color{blue}[DJW] : #1}\fi}

\hypersetup{
  colorlinks,
  linkcolor={red!70!black},  
  citecolor={red!70!black},
  urlcolor={blue!70!black}
}

\definecolor{myforestgreen}{RGB}{34,139,34} 

\newcommand{\header}[1]{\paragraph{\textnormal{\textbf{#1}}}}

\renewenvironment{quote}
  {\list{}{\rightmargin=.5cm \leftmargin=.5cm}%
   \item\relax}
  {\endlist}

\definecolor{lightred}{RGB}{255, 200, 200}
\definecolor{lightblue}{RGB}{200, 220, 255}
\definecolor{lightgrey}{RGB}{240, 240, 240}

\newcommand{\hlc}[2][yellow]{{%
    \colorlet{foo}{#1}%
    \sethlcolor{foo}\hl{#2}}%
}

\newcolumntype{C}[1]{>{\centering\arraybackslash}p{#1}}

\newcommand{\otrace}{\textsc{OTrace}\ }

\begin{document}
\sloppy

\title{Data Traceability for Privacy Alignment}
\titlenote{Work-in-progress paper at the 4th ACM Symposium on Computer Science and Law, March 25--27, 2025, Munich, Germany}

\author{Kevin Liao}
\affiliation{%
    \institution{MIT and Harvard Law School}}
\email{kevliao@mit.edu}

\author{Shreya Thipireddy}
\affiliation{%
    \institution{MIT}}
\email{shreyat@mit.edu}

\author{Daniel Weitzner}
\affiliation{%
    \institution{MIT}}
\email{weitzner@mit.edu}

\begin{abstract}

This paper offers a new privacy approach for the growing ecosystem of services---ranging from open banking to healthcare---dependent on sensitive personal data sharing between individuals and third-parties. While these services offer significant benefits, individuals want control over their data, transparency regarding how their data is used, and accountability from third-parties for misuse. However, existing legal and technical mechanisms are inadequate for supporting these needs. A comprehensive approach to the modern privacy challenges of accountable third-party data sharing requires a closer \emph{alignment} of technical system architecture and legal institutional design.

In order to achieve this privacy alignment, we extend traditional security threat modeling and analysis to encompass a broader range of privacy notions than has been typically considered. In particular, we introduce the concept of \emph{covert-accountability}, which addresses the risk from adversaries that may act dishonestly but nevertheless face potential identification and legal consequences. As a concrete instance of this design approach, we present the \otrace protocol, designed to provide traceable, accountable, consumer-control in third-party data sharing ecosystems. \otrace empowers consumers with the knowledge of who has their data, what it is being used for, what consent or other legal terms apply, and whom it is being shared with. By applying our alignment framework, we demonstrate that \otrace's technical affordances can provide more confident, scalable regulatory oversight when combined with complementary legal mechanisms. 
\end{abstract}

\maketitle

\section{Introduction}
\label{sec:introduction}

\epigraph{``Publicity is justly commended as a remedy for social and industrial diseases. Sunlight is said to be the best of disinfectants; electric light the most efficient policeman''~\cite{brandeis_1914}.}{\textit{Louis Brandeis, then-U.S. Supreme Court Justice}}


The first several decades of the online privacy debate has been largely focused on a single problem: how to reconcile fundamental privacy rights with the explosion of the surreptitious tracking and profiling that enables the advertising and marketing driven Internet economy. The results have fallen far short of Justice Brandeis' call for sunlight~\cite{zuboff2023age}. With the unfulfilled privacy promises in full view, we turn our attention here to a new set of privacy challenges that we believe will define the next era of privacy challenges, what we call ``modern privacy.'' All around us there is a new class of services---ranging from open banking, to personal fitness, expanded healthcare, and employment---using sensitive personal data. These services depend on individuals choosing to share information over which they do have control with a growing range of third-parties. New services have already shown the potential to provide benefit to individuals as well as innovation opportunities in the marketplace. While individuals want to share personal data, and want to get some value or insight from doing so, they do not want to simply abandon their data to the whims of the third-parties who get hold of it. Rather, individuals want to be able to control how their data is used, keep track of with whom it is shared, and hold these third-parties accountable for misuse, whether because of breaking the terms of a consent agreement or violating a law, regulation or contract. 

Today, neither the technical infrastructure nor legal and organizational mechanisms exist to build a data ecosystem able to support the control and accountability needs of consumers or companies in these new service contexts.  We show that a comprehensive approach to the modern privacy challenges of third-party data sharing requires a closer \emph{alignment} of technical system architecture and legal institutional design.

Through most of the history of contemporary privacy, technical systems and legal architecture for privacy protection has been developed on largely independent tracks. In computer science, we see sustained work on access control, private computation, deidentification and information leakage such as differential privacy. Yet, there is comparatively little attention to purpose specification and misuse detection. The results leave substantial gaps if we are to support robust, scalable privacy protection in the modern world of third-party data sharing that is upon us. To begin with, companies are not sure how to comply with privacy laws and regulations without incurring undue costs. As an example, a recent leaked memo from Facebook reported that it would take five years and hundreds of person-years of coding resources to be able to comprehensively track internal use of personal data in the Facebook infrastructure alone ~\cite{facebook-leak}. And beyond this, companies lack assurance that third-parties are honoring their contractual data obligations~\cite{contractual-obligations}.

One consequence of the lack of personal data control in enterprise systems is the corresponding inability of regulators to enforce privacy laws already on the books~\cite{kerry2018protecting}. And the end result of uncontrolled handling of personal data by companies followed by lack of confident, scalable enforcement be regulators is a justifiable lack of trust by users~\cite{chen2021battle}. So, while we have significant and growing privacy regulation on the books, all of the technical infrastructure to support compliance has fallen far behind. This is a gap that must be addressed before we can see fully trusted, accountable modern third-party data sharing services be deployed.

In order to assess the alignment between legal and technical affordances in any system, we extend traditional security threat modeling and analysis frameworks to encompass a broader range of privacy notions than has been typically considered. Specifically, we introduce a new security guarantee, \emph{covert-accountability}, which considers a covert adversary that may act arbitrarily but faces the risk of identification and legal consequences for dishonest behavior. This expanded framework can be used to analyze systems in which privacy properties can be relied upon only with a specified combination of legal and technical properties.

As a concrete instance of this design approach, we present the \otrace protocol, designed to provide traceable, accountable, consumer-controllable data sharing in $n$-party ecosystems. \otrace empowers consumers with the knowledge of where their data is, who has it, what it is being used for, and whom it is being shared with. By applying our new threat model analysis framework to \otrace, we clarify the contours of where technical guarantees end and legal guarantees begin, helping to determine how legal frameworks can align with \otrace's design to achieve the desired privacy outcomes. Legal frameworks here include statutes, common law rules, administrative regulations or legally-enforceable rules contained in contracts. Not only do we find that existing consumer protection frameworks can provide legal guarantees where \otrace's technical safeguards leave off, but we also demonstrate that \otrace's technical affordances can provide more confident, scalable regulatory oversight.

As a result, when law and technology are aligned for traceability, companies will have an easier time managing personal data according to its intended uses, regulators will be able to detect violates of privacy rules at scale, and individuals will have a greater sense of trust that their personal data is being used as they expect and deleted or corrected when they desire.

In sum, the main contributions of this paper are:
\begin{itemize}
    \item an analysis of legal and technical approaches to privacy compliance, highlighting the gaps they leave and the need for new ways of thinking (Section~\ref{sec:background}),
    \item the introduction of a legal-technical alignment framework that integrates reasoning not only about a protocol's technical guarantees but, for the first time, its legal guarantees as well (Section~\ref{sec:framework}),
    \item the design of \otrace, a data traceability protocol that empowers consumers to track the use of their data, even across company borders (Section~\ref{sec:protocol}), and
    \item the design of a legal and regulatory framework surrounding data traceability to hold companies accountable for data misuse (Section~\ref{sec:accountability}).
\end{itemize}

\section{Privacy Law and Technology}
\label{sec:background}

In this section, we analyze legal and technical approaches to privacy compliance, highlighting the gaps they leave and the need for new ways of thinking.

\subsection{Legal Frameworks}

\header{US federal enforcement}
The Federal Trade Commission (FTC) serves as the primary privacy regulator in the US.\footnote{Apart from the FTC, other sector-specific regulators also play a role in overseeing privacy protections, such as the Consumer Financial Protection Bureau (CFPB)~\cite{cfpb} for financial data and the Department of Health and Human Services (HHS)~\cite{hhs} for health data.} In addition to its specific statutory mandates, the FTC exercises broad enforcement authority under Section 5 of the FTC Act, which prohibits ``unfair or deceptive acts or practices''~\cite{ftc_section_5}. Though not all-encompassing, this expansive mandate allows the FTC to address a wide range of consumer privacy issues---any action that deceives or harms consumers arguably falls within the scope of its Section 5 jurisdiction.

Instead of litigating Section 5 cases to trial, the FTC typically negotiates settlements with companies, resulting in \emph{consent decrees}~\cite{solove2014ftc}. These settlements commonly require the company to cease the deceptive or unfair practices outlined in the complaint, avoid future Section 5 violations, and establish a 20-year monitoring program with regular reporting to the FTC to ensure compliance.

While the FTC is rightly credited for shaping much of the common law of consumer privacy in the US~\cite{solove2014ftc}, its efforts have struggled to keep pace with the growing complexity and volume of data flows. Consider the infamous Facebook-Cambridge Analytica scandal~\cite{cambridge-analytica}. In 2014, due to the extremely porous nature of the Facebook developer APIs and lack of any controls on outward data flow to third-parties, data from over 87 million Facebook profiles was harvested by Cambridge Analytica, a UK-based political consulting firm ~\cite{weitzner_how_2023}. Cambridge Analytica then went on to use that data for commercial, political advertising without the consent or even notice of those users.

Remarkably, this all occurred while Facebook was already under a 2011 FTC consent decree following an earlier privacy violation~\cite{facebook2011consent}. The extant decree required Facebook to  implement a ``comprehensive privacy program'' and undergo regular third-party audits for twenty years, yet these glaring violations slipped through the cracks of both Facebook's internal compliance program and, more importantly, went unnoticed by the FTC-order audits ~\cite{vladeck_facebook_nodate} .


\header{US state and private enforcement}
State consumer privacy laws and privacy enforcement by state attorneys general (AGs) have increasingly stepped in to the fill regulatory gap left by the lack of a comprehensive federal consumer privacy framework~\cite{citron2016privacy}. California paved the way with the California Consumer Privacy Act  of 2018 (CCPA)~\cite{california2018privacy}, the first omnibus data protection law in the US. Following in the footsteps of the EU's General Data Protection Regulation (GDPR)~\cite{EuropeanParliament2016a}, the CCPA recognized various data rights, including the rights to access, correct, and delete personal information. Since then, nineteen states have enacted similar laws~\cite{state-privacy-tracker}. 


Privacy compliance is a rapidly moving target at state and private levels. While state-level enforcement and new private rights of action can complement federal privacy protection efforts, they also introduce complexity for companies navigating the requirements of multiple jurisdictions. As Citron notes, ``offices of state attorneys general [are] laboratories of privacy enforcement''~\cite{citron2016privacy}, but the growing variations among state laws risk creating a regulatory patchwork that complicates compliance. Moreover, current enterprise data systems lack the agility to keep pace with these diverse and shifting requirements.




\header{EU privacy law}
Adopted in 2016, the GDPR~\cite{EuropeanParliament2016a} has fundamentally changed how companies operating within the EU handle personal data. For many firms offering services globally, the GDPR has become the de facto set of rules governing how they handler personal data, regardless of whether the data subjects of EU citizens or not. Unlike in the US, where companies can process personal data unless explicitly restricted by law, the GDPR mandates that personal data can only be processed with a valid ``legal basis.'' This requirement to maintain control over personal data at all times introduces several key obligations for companies. First, they must track and enforce the legal basis for processing personal data within their systems. Moreover, they must be equipped to honor data subjects' requests, such as those invoking the ``right to be forgotten.'' Meeting these requirements necessitates technological adaptations to manage the legal constraints imposed by the GDPR. Unfortunately, enterprise data systems currently lack the capabilities to comply with these regulations effectively~\cite{DBLP:conf/vldb/KraskaSBSW19}.

\subsection{Technical Approaches}

\header{Privacy-compliance tools}
On the technical side, there is a rich body of academic literature on privacy-enhancing technologies~\cite{heurix2015taxonomy}. We focus here on the subset of these technologies tha are most relevant, privacy-compliance tools, which are designed to facilitate adherence to privacy laws and regulations. We categorize these tools into three broad categories:
\begin{itemize}
    \item \emph{violation-prevention tools}, designed to prevent policy violations (e.g., purpose limitations)~\cite{DBLP:conf/sp/SenGDRTW14, DBLP:conf/uss/WangKNPSS0LS22, DBLP:conf/vldb/KraskaSBSW19, DBLP:conf/nsdi/WangKM19, DBLP:conf/sp/FerreiraBSS23, DBLP:conf/osdi/BurkhalterKVSH21};
    \item \emph{rights-automation tools}, designed to automate processing in response to data rights requests (e.g., data deletion requests)~\cite{DBLP:conf/uss/Cohn-GordonDNCR20, edna, fides, k9db, DBLP:conf/sp/BourtouleCCJTZL21}; and
    \item \emph{rights-interoperability tools}, designed to standardize interactions between parties in the handling of data rights~\cite{dti, drp, gpc}.
\end{itemize}

Violation-prevention tools focus on preventing policy violations before they occur, rather than addressing them after the fact. However, the sheer scale and complexity of today's data ecosystems have revealed the limitations of preventative tools~\cite{DBLP:journals/cacm/WeitznerABFHS08, DBLP:journals/ftsec/FeigenbaumJW20}. Indeed, these tools can provide strong technical guarantees by leveraging techniques such as code analysis~\cite{DBLP:conf/sp/SenGDRTW14, DBLP:conf/uss/WangKNPSS0LS22, DBLP:conf/nsdi/WangKM19, DBLP:conf/sp/FerreiraBSS23}, database design~\cite{DBLP:conf/vldb/KraskaSBSW19}, trusted hardware~\cite{DBLP:conf/uss/WangKNPSS0LS22, DBLP:conf/nsdi/WangKM19}, and cryptographic protocols~\cite{DBLP:conf/osdi/BurkhalterKVSH21}. However, these approaches significant barriers to practical adoption, since many require developers to manually annotate data, queries, or libraries~\cite{DBLP:conf/sp/SenGDRTW14, DBLP:conf/uss/WangKNPSS0LS22, DBLP:conf/vldb/KraskaSBSW19, DBLP:conf/sp/FerreiraBSS23}, while others necessitate the use of specialized hardware~\cite{DBLP:conf/uss/WangKNPSS0LS22, DBLP:conf/nsdi/WangKM19} or computing infrastructure~\cite{DBLP:conf/vldb/KraskaSBSW19, DBLP:conf/nsdi/WangKM19, DBLP:conf/osdi/BurkhalterKVSH21}.

Rights-automation tools, by contrast, aim to reduce the costs of complying with data rights requests. Like violation-prevention tools, many rights-automation tools require data annotations~\cite{DBLP:conf/uss/Cohn-GordonDNCR20, edna, fides, k9db} or specialized infrastructure~\cite{DBLP:conf/uss/Cohn-GordonDNCR20, edna, fides, k9db}, which limits their practical adoption. A notable exception is DELF~\cite{DBLP:conf/uss/Cohn-GordonDNCR20}, which has been successfully deployed at Facebook. However, its implementation required efforts that only a large company like Facebook could afford and is highly specialized to its data systems. Unfortunately, none of these tools provide consumers assurance that their rights have actually been fulfilled.

Likewise, rights-interoperability tools are designed to alleviate the technical and administrative burdens companies face when handling consumer data rights requests. Tools such as the Data Transfer Initiative (DTI)~\cite{dti}, the Data Rights Protocol (DRP)~\cite{drp}, and Global Privacy Control (GPC)~\cite{gpc} offer APIs that facilitate implementation and interoperability, encouraging broader adoption. However, as with rights-automation tools, none of these provide consumers a guarantee that their rights have been fully honored. GPC, a browser signal that communicates user preferences, such as opting out of data tracking, ``may become a legally binding opt-out signal in California,'' though this remains untested~\cite{gpc-faq}.

\header{Accountable systems}
A rich body of literature explores accountability through a computer science lens, e.g., \cite{DBLP:journals/cacm/WeitznerABFHS08, DBLP:journals/ftsec/FeigenbaumJW20, DBLP:conf/uss/FranklePSGW18, DBLP:conf/socialcom/PatoPJSW11, kroll2014secure, kroll165accountable, DBLP:conf/fat/Kroll21, DBLP:conf/icnp/ArgyrakiMIAS07, DBLP:conf/ndss/BackesDHU09, DBLP:conf/policy/BreauxAKK06, DBLP:journals/compsec/BreauxAS09, DBLP:conf/re/BreauxVA06, DBLP:conf/scn/CamenischHL06, DBLP:conf/icdcit/Datta14, vedder2017accountability, DBLP:conf/osdi/HaeberlenARD10, DBLP:conf/sosp/HaeberlenKD07, DBLP:conf/esorics/JagadeesanJPR09, DBLP:conf/ccs/KustersTV10, DBLP:conf/fat/YoungRKSSWH19, DBLP:conf/nspw/FeigenbaumJW11, DBLP:conf/ctrsa/Lindell08a}. 
Broadly speaking, accountable systems are information systems in which ``policy violations are punished'' with some degree of certainty~\cite{DBLP:journals/ftsec/FeigenbaumJW20}. These systems contrast with, yet complement, violation-prevention tools that aim to stop policy violations before they occur. This body of work can generally be categorized along two dimensions: (1) their design focus, whether legal or technical, and (2) the type of accountability guarantees they aim to achieve, again either legal or technical. Despite their shared focus on technical design, many of these works emphasize technical accountability, while others lend themselves to legal accountability, but without significantly developing the underlying legal frameworks.

Unfortunately, much like technical guarantees from violation-prevention tools, technical accountability guarantees are often limited in their applicability and costly to implement. While legal accountability guarantees can help address these limitations, technical design alone is often a crude tool for delivering legal accountability, especially when the underlying legal frameworks are underdeveloped.

\header{Formal privacy analysis frameworks}

Provable security is the technique of formally proving a protocol's security guarantees, which includes certain classes of privacy guarantees, notably secrecy~\cite{goldwasser1984probabilistic,goldreich1987play}.\footnote{For an in-depth treatment on provable security that goes beyond this brief overview, see Lindell's tutorial~\cite{lindell2017simulate}.} Unlike earlier heuristic approaches, provable security allows for guarantees that can be quantified with probabilistic precision. A key step in the analysis is specifying the \emph{threat model}, which defines the adversary's capabilities that the protocol must withstand to ensure the desired guarantees.

A baseline threat model is the \emph{passive} (or \emph{semi-honest}) adversary model. In this model, the adversary controls one or more parties and follows the protocol exactly as prescribed, but it tries to learn more information than permitted during the protocol's execution. A \emph{passive-secure} protocol ensures no inadvertent information leakage so long as all parties adhere to the protocol. However, the passive adversary model rests on a tenuous assumption, making it a relatively weak threat model with limited security guarantees. Security can be compromised if an adversary deviates from the protocol, such as by sending an arbitrary message out of turn. In practice, a motivated adversary is unlikely to remain passive and will exploit any available means to break the protocol. Nonetheless, while passive security may be too weak to translate into meaningful, real-world guarantees, passive-secure protocols serve as valuable stepping stones toward securing against stronger adversaries.

A more robust threat model is the \emph{active} (or \emph{malicious}) adversary model, where the adversary may behave arbitrarily. Accordingly, \emph{active-secure} protocols maintain their security guarantees even when parties deviate from the protocol. These protocols are better suited for environments where parties cannot be trusted to behave honestly. However, achieving security against such a strong adversary often requires heavyweight cryptographic tools, such as multi-party computation~\cite{goldreich1998secure} and zero-knowledge proofs~\cite{goldwasser1985knowledge}, or trusted hardware, such as Intel SGX~\cite{costan2016intel}, which makes practical deployment challenging and costly. Violation-prevention tools, for example, typically operate within this threat model.

The \emph{covert} adversary model~\cite{aumann2010security} aims to strike a balance between the simplicity and efficiency of protocols realizable in the passive model and the robustness of guarantees realizable in the active model. As in the active model, an adversary may deviate arbitrarily from the protocol, but a \emph{covert-secure} protocol ensures that any deviation by a malicious party will be detected by honest parties with high probability. This makes covert-secure protocols suitable for certain commercial, political, or social contexts where the risk of detection is a sufficient deterrent to malicious behavior, e.g., news of malicious behavior could lead to significant reputational harm.

These traditional threat models for privacy guarantees are often limited in scope, focusing narrowly on secrecy---whether sensitive information has been leaked. However, real-world privacy threats are more complex, encompassing issues such as unauthorized data use, exceeding permitted purposes, improper sharing, and prolonged retention beyond agreed-upon limits. Effectively addressing these challenges requires an expanded analytical framework that considers both legal and technical guarantees.

\subsection{The Need to Bridge Law and Technology}

Regulators need new tools to scale their oversight, while companies need new tools to scale their compliance efforts. Although privacy-compliance tools and accountable systems offer some solutions, violation-prevention and technical accountability tools face significant barriers to adoption, rights-automation and rights-interoperability tools fail to instill consumer confidence that their rights are being upheld, and legal accountability tools have yet to develop the supporting legal frameworks. Moreover, technical frameworks for privacy analysis have been used effectively for secrecy guarantees, but not the broader set of privacy guarantees contemplated by privacy law. All of this highlights the need to rethink how law and technology can come together to provide comprehensive privacy guarantees across both domains.

%
%
%
%
%
%
%
%
%
%
%
%
%
%
%
%
%


%

\section{A Legal-Technical Privacy Alignment Framework}
\label{sec:framework}

We first motivate and preview our new legal-technical alignment framework that enables reasoning not only about a system's technical guarantees but, for the first time, its legal guarantees as well. This framework helps delineate boundaries between the two, clarifying where technical guarantees end, where legal guarantees begin, and how they can mutually reinforce each other. Ultimately, this framework facilitates the design of practical protocols with both legal and technical guarantees.





\header{The inadequacy of traditional technical approaches to privacy} 
Traditional technical approaches to information privacy have long prioritized secrecy~\cite{DBLP:journals/cacm/WeitznerABFHS08}. If we cannot trust companies to handle personal data responsibly, the most effective safeguard might be to prevent them from accessing it altogether. Consequently, the privacy community has developed a range of cryptographic techniques that enable data processing without exposing the underlying data. For example, homomorphic encryption~\cite{DBLP:journals/csur/AcarAUC18} allows computations to be performed on encrypted data, ensuring that the underlying raw data remains inaccessible. Secure multiparty computation~\cite{DBLP:journals/cacm/Lindell21} enables multiple parties to jointly compute a function over their private inputs while revealing only the final result. Federated learning~\cite{DBLP:journals/ftml/KairouzMABBBBCC21} facilitates machine learning model training across decentralized devices, allowing participants to contribute to model updates without sharing raw data. Differential privacy~\cite{DBLP:conf/tamc/Dwork08} provides another mechanism for privacy-preserving data analysis, introducing statistical noise to protect individual contributions while still allowing for meaningful aggregate insights.

Yet, in today's information-sharing ecosystem, this limited approach to privacy fails to address the modern privacy needs we have identified. In modern privacy contexts consumers actually intend to share their personal data but they object to having that information used in ways that subvert their expectations or work against their interests. For example, a user might be comfortable sharing their location data with a navigation app to receive real-time traffic updates but would object if that same data were later sold to an advertiser without their knowledge. Similarly, individuals may accept data collection in the context of improving product recommendations but oppose its use in ways that manipulate or discriminate against them.

As such, privacy in real-world contexts is inherently more complex than the limited notion of secrecy. Traditional secrecy-based models assume that privacy threats arise primarily from unintended access to sensitive data, but many contemporary concerns stem from the ways in which data is used rather than merely accessed. Modern privacy threats encompass issues such as unauthorized data use, exceeding permitted purposes, improper sharing, and prolonged retention beyond agreed-upon limits. Secrecy is not always necessary nor is it ever sufficient for addressing these concerns; it necessitates mechanisms for ensuring that data is processed in accordance with user expectations and legal obligations.

Some technical approaches attempt to enforce these kinds of privacy beyond secrecy, such as requiring companies to use specialized software or hardware to ensure that data can only be processed according to user-permitted purposes~\cite{DBLP:conf/uss/WangKNPSS0LS22, DBLP:conf/nsdi/WangKM19, DBLP:conf/vldb/KraskaSBSW19, DBLP:conf/nsdi/WangKM19, DBLP:conf/osdi/BurkhalterKVSH21}. Trusted execution environments, for instance, can provide secure enclaves where computations are executed in a verifiable manner, ensuring that data remains protected even from the infrastructure processing it~\cite{DBLP:conf/uss/WangKNPSS0LS22, DBLP:conf/nsdi/WangKM19}. However, these approaches remain computationally expensive and introduce complex implementation requirements.

Such privacy threats do not align well with traditional technical approaches to privacy because they are governed less by technical constraints and more by legal, ethical, and contextual considerations. The challenge is not merely preventing adversaries from accessing raw data but ensuring that data is used in ways that align with individual rights and expectations. Nevertheless, we would still like to be able to evaluate privacy protocols with the same rigor as with traditional approaches wherever possible while incorporating an ``escape hatch'' for cases where alternative methods, namely legal accountability, are more appropriate.

\header{The covert-accountability guarantee}
Our approach extends the covert adversary model by introducing a new \emph{covert-accountability} guarantee alongside the traditional, probabilistic guarantees. This means that if a malicious party deviates from the protocol, they face an increased risk of identification and legal consequences. The deterrent effect of legal accountability is thus driven by both the likelihood and severity of those consequences.

While legal accountability may lack the mathematical rigor and precision of traditional guarantees, it has long been an effective means of governing societal behavior. As Weitzner et al.\ \cite{DBLP:journals/cacm/WeitznerABFHS08} explain in their work on information accountability:

\begin{quote}
``The vast majority of legal and social rules that form the fabric of our societies are not enforced perfectly or automatically, yet somehow most of us still manage to follow most of them most of the time. We do so because social systems built up over thousands of years encourage us, often making compliance easier than violation. For those rare cases where rules are broken, we are all aware that we may be held accountable through a process that looks back through the records of our actions and assesses them against the rules\ldots The information-accountability framework more closely mirrors the relationship between the law and human behavior than do the various efforts to enforce policy compliance through access control over information.''
\end{quote}
Thus, a key objective in the technical design of accountability-based protocols is to make ``bad acts visible to all concerned''~\cite{DBLP:journals/cacm/WeitznerABFHS08}, thereby increasing the likelihood of corrective action~\cite{DBLP:conf/nspw/FeigenbaumJW11}.

\header{Mixing adversary models and guarantees}
We also adopt a mixed adversary model to account for differing levels of trustworthiness among parties in commercial settings. That is, some parties may be fully trusted or passive, serving as roots of trust, while others may deviate arbitrarily. Moreover, protocols analyzed under our framework can exhibit a combination of different types of guarantees. Traditionally, protocols are designed to be uniformly secure across all guarantees. In contrast, our framework allows protocols to exhibit, for example, covert-security in some respects and covert-accountability in others.

\paragraph{\textnormal{The result is a legal-technical alignment framework that not only clarifies the boundaries between legal and technical guarantees but also informs how they can mutually reinforce each other. For example, protocols designed within this framework can offer robust active-security or covert-security guarantees where practicable, while relying on covert-accountability guarantees in contexts that would previously require heavyweight technical tools. The technical design helps alleviate burdens of legal enforcement while also enhancing its effectiveness and scalability; the legal design delivers practical guarantees without the need for impractical technical tools. Seeing the framework in action over the next sections will allow one to more fully appreciate its benefits.}}
\section{Data Traceability}
\label{sec:protocol}

To bridge the gap between privacy law and technology, we propose a simple yet powerful idea: \emph{data traceability}. At its core, data traceability aims to enhance transparency by enabling the tracking of who has access to data and how it is being used---especially as it moves across company boundaries~\cite{DBLP:conf/wpes/PullsPW13}. Traceability addresses several key information accountability challenges:
\begin{itemize}
    \item Within organizations, it helps companies manage personal information more effectively, especially across siloed teams, facilitating compliance with privacy laws and regulations. 
    \item Between companies, it helps hold data-sharing partners accountable for their contractual obligations. 
    \item For regulators, traceability provides visibility into how companies handle personal information (or at least how they claim to), enabling scalable privacy enforcement. 
    \item For consumers, it provides trust through visibility into how their data is used and empowers them with meaningful control, enabling them to align data practices with their expectations and preferences.
\end{itemize}
In this section, we outline our technical design for traceability.


\subsection{Overview}

\header{Roles} Traceability relies on a well-defined structure of participants in a data-sharing ecosystem. Broadly, these participants fall into three primary roles:
\begin{itemize}
    \item \emph{Consumers.} Individuals who share their data in exchange for useful services.
    \item \emph{Providers.} Companies that collect, process, and \emph{provide} consumer data.
    \item \emph{Recipients.} Companies that \emph{receive} and process consumer data to deliver those useful services.
\end{itemize}

To illustrate how these roles interact in practice, consider a running example from open banking~\cite{jeng2022open}. Suppose a consumer, Alice, wants to use a personal finance management application, \emph{MoneyApp}, which aggregates and analyzes financial information from multiple accounts. To do so, she initiates a data-sharing relationship in which her bank, \emph{FirstBank}, which holds her account and transaction information, serves as a provider, while \emph{MoneyApp} serves as the recipient, accessing and processing the information to deliver personalized financial insights.

However, Alice is not limited to just one such data-sharing relationship. She may also use health applications that pull data from her medical records, smart home platforms that process energy consumption patterns, or insurance services that access her driving history. As the number of these relationships grows, keeping track of who has access to her data and ensuring it is used appropriately can become increasingly complex and difficult to manage.  To address this challenge, traceability introduces an additional role:
\begin{itemize}
    \item \emph{Traceability agents.} Consumer agents that enable consumers to track who has their data, how it is being used, and whether its use aligns with their expectations.
\end{itemize}

\header{Concepts}
With these high-level roles in mind, we now turn to explaining how traceability works in practice, and ultimately how it will be implemented in code, through bite-sized components called \emph{concepts}, each serving as a ``self-contained unit of functionality''~\cite{jackson2021essence}. Concept design is a new software engineering approach to organize the functionality that computers provide into independent and reusable units. The concepts of a software system are at the same time the ideas you need to understand in order to use it and the implemented services that the system provides.  The lack of coherent concept design for privacy-relevant systems in consumer services lead to both confusion on the part of consumers, thwarting user control, as well as inability for enterprise system designers to handle data in a respectful manner across complicated data ecosystems. By using the approach of concept design to express each concept in terms of its abstract observable behavior, we are able to produce compelling explanations of privacy-relevant behavior for all stakeholders that are independent of any particular user interface. At the same time, we can separate the essence of the scheme from its implementation as services or network protocols.

Our traceability concepts are fully detailed in Appendix~\ref{sec:concepts}. Here, we provide a high-level overview to convey their core purpose and how they fit together.

\textsc{Consent}. One of the cornerstone principles in privacy law---and, by extension, a fundamental concept in traceability---is \textsc{Consent}. Privacy consent represents an agreement between a consumer and a data controller regarding what data the company may collect, how it may be used, whether it may be shared or sold, and so on. For example, in the open banking scenario, Alice has separate consent agreements with \emph{FirstBank} and \emph{MoneyApp}. These agreements, likely established when she first signed up for each service, define how each party may handle her financial data.

Our \textsc{Consent} concept incorporates several key characteristics:
\begin{itemize}
    \item \emph{Standardized.} Consent should be structured in a clear and consistent format, allowing consumers to easily understand their agreements across different services. Standardization also facilitates interoperability between providers, recipients, and traceability agents. Examples of recent formalized consent ontologies include the Visa Consent Management Framework~\cite{visa-cmf} and the Fides Privacy Engineering and Compliance Framework~\cite{fides_taxonomy}.
    \item \emph{Granular.} Consumers should have the ability to grant or deny consent at a detailed level. Rather than providing broad, all-or-nothing approval, they should be able to specify which types of data can be collected, how it can be used, and under what circumstances it can be shared.
    \item \emph{Revocable.} Consent should not be a one-time decision. Consumers must have the ability to withdraw or modify their consent at any time, ensuring they retain ongoing control over their personal data as their preferences, needs, or circumstances change.
\end{itemize}

\textsc{RightsRequest.} Another key privacy control recognized by many privacy laws is a set of individual privacy rights, which we capture under our \textsc{RightsRequest} concept. Many privacy laws, including the GDPR and CCPA, grant consumers a range of rights, including the right to access, correct, delete their personal information held by a company. For example, if Alice is a California resident and \emph{MoneyApp} is a covered entity under the CCPA, she may request that \emph{MoneyApp} delete her personal data. \emph{MoneyApp} is then obligated to comply with her request, subject to certain exemptions.

\textsc{Authorization.} Every data-sharing relationship begins with an \textsc{Authorization}, in which a consumer grants a recipient limited access to data about them held by a provider. For example, Alice authorizes \emph{MoneyApp} to access her financial data stored by \emph{FirstBank}.

\textsc{Introduction.} Every traceability relationship begins with an \textsc{Introduction}, in which a consumer connects one party (e.g., a provider or recipient) to another party (e.g., their traceability agent). For example, while Alice authorizes the data-sharing relationship between \emph{MoneyApp} and \emph{FirstBank}, she may also introduce each party (if she has not done so already) to her traceability agent, allowing it to track how \emph{MoneyApp} and \emph{FirstBank} share and use her data.

\textsc{Attestation.}
The fundamental building block of traceability is the \textsc{Attestation}. When a party (whether a provider or a recipient) performs an action (e.g., collection, processing, sharing) on consumer data, the controller ``goes on the record'' by submitting an attestation of that action to the consumer's traceability agent. This attestation is an affirmative declaration by the controller that it has carried out the specified action on the consumer's data. Collectively, these attestations create a paper trail of the controller's data activities.

The purposes of these attestations are at least twofold. First, they provide consumers transparency into how controllers are handling their data, within and across data controllers. Second, they serve as ``hooks'' to hold controllers accountable in the event of data misuse. That is, if a controller's actual practices diverge from their attested actions, these discrepancies can substantiate claims of non-compliance or deceptive practices in future enforcement actions.

To capture various kinds of data events, our traceability protocol distinguishes between different types attestations:
\begin{itemize}
    \item \emph{consent attestations} document consent agreements between consumers and controllers, including the types of data covered, purpose limitations, retention duration, and sharing conditions;
    \item \emph{sharing attestations} document data sharing agreements between providers and recipients, including the types of data covered, purpose limitations, retention duration, and resharing conditions;
    \item \emph{access attestations} document data-access actions between recipients and providers, including the types of data shared, purposes of access, and timestamps of these actions;
    \item \emph{process attestations} document data-process actions local to a controller, including the types of data used, purposes of use, and timestamps of these actions; and
    \item \emph{request attestations} document compliance actions by controllers in response to data rights requests, such as data deletion or correction requests.
\end{itemize}


\begin{figure*}[t]
    \centering
    \includegraphics[width=0.75\textwidth]{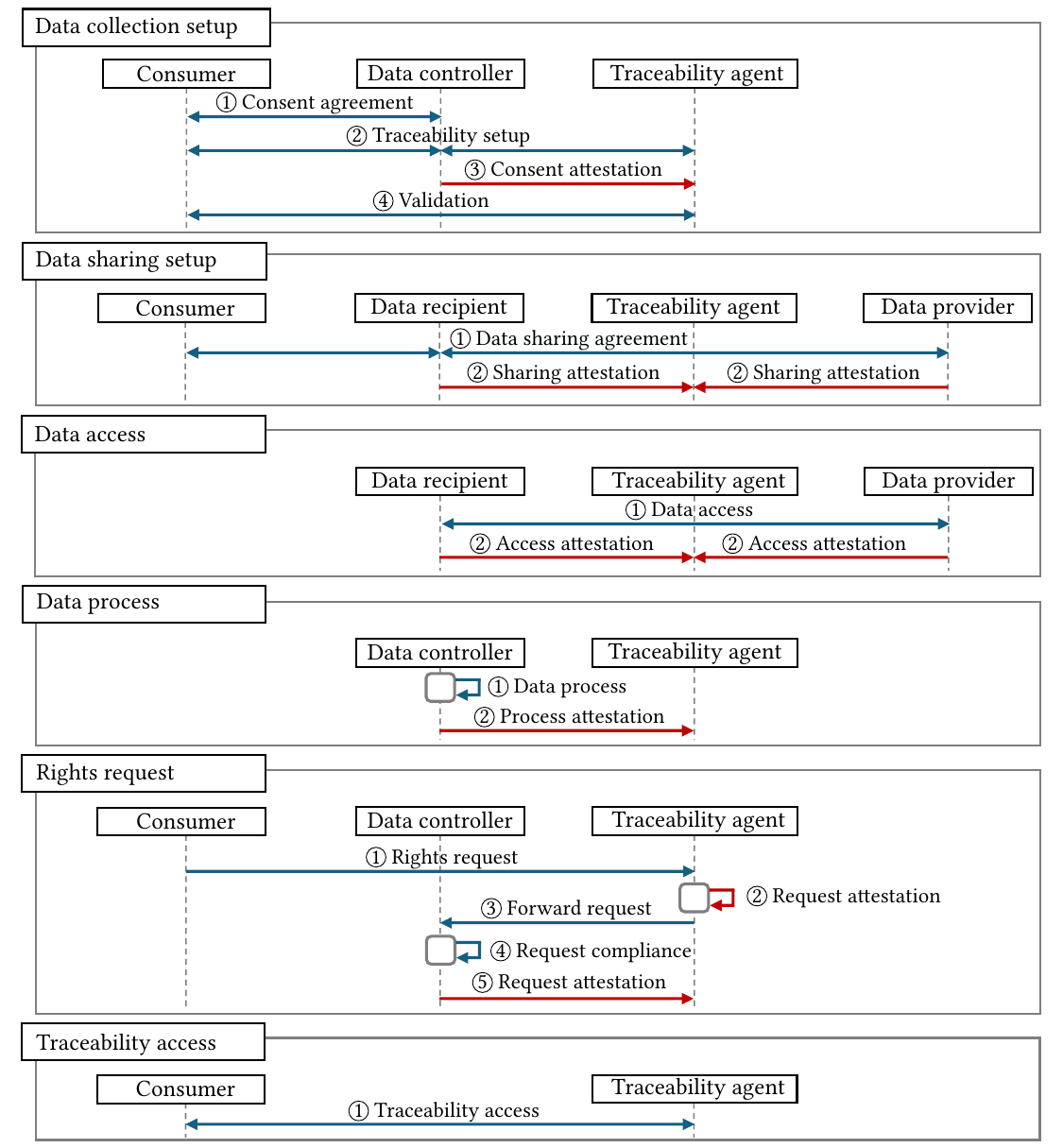}
    \caption{High-level overview of the \otrace traceability protocol.}
    \label{fig:traceability}
\end{figure*}

\begin{table*}[t]
    \centering
    \begin{tabular}{|c|c|C{3.1cm}|C{3.1cm}|C{3.1cm}|C{3.1cm}|}
    \hhline{~~|----}
    \multicolumn{1}{c}{} & \multicolumn{1}{c|}{} & \multicolumn{4}{c|}{\textbf{Data controller (threat model)}}\\
    \hhline{~~|----}
    \multicolumn{1}{c}{} & \multicolumn{1}{c|}{} & \cellcolor{lightblue}\textbf{Provider} & \cellcolor{lightblue}\textbf{Recipient} & \cellcolor{lightred}\textbf{Provider} & \cellcolor{lightred}\textbf{Recipient} \\
    \multicolumn{1}{c}{} & \multicolumn{1}{c|}{} & \cellcolor{lightblue}\textbf{(trusted/semi-honest)} & \cellcolor{lightblue}\textbf{(covert)} & \cellcolor{lightred}\textbf{(covert)} & \cellcolor{lightred}\textbf{(covert)} \\
    \hhline{*{6}-}
    & \textbf{Consent}  & \cellcolor{lightblue} assumed & \cellcolor{lightblue} consumer-dependent & \cellcolor{lightred} consumer-dependent & \cellcolor{lightred} consumer-dependent \\
    \hhline{|~|*{5}-}    
    & \textbf{Sharing} & \cellcolor{lightblue} assumed & \cellcolor{lightblue} covert-secure & \cellcolor{lightred} covert-accountable & \cellcolor{lightred} covert-accountable \\
    \hhline{|~|*{5}-}    
    & \textbf{Access} & \cellcolor{lightblue} assumed & \cellcolor{lightblue} covert-secure & \cellcolor{lightred} covert-accountable & \cellcolor{lightred} covert-accountable \\
    \hhline{|~|*{5}-}    
    & \textbf{Process} & \cellcolor{lightblue} assumed & \cellcolor{lightblue} covert-accountable & \cellcolor{lightred} covert-accountable & \cellcolor{lightred} covert-accountable \\
    \hhline{|~|*{5}-}    
    \multirow{-5}{*}[0em]{\rotatebox[origin=c]{90}{\textbf{Attestation}}} & \textbf{Request} & \cellcolor{lightblue} assumed & \cellcolor{lightblue} covert-secure & \cellcolor{lightred} covert-accountable & \cellcolor{lightred} covert-accountable \\
    \hline
    \end{tabular}
    \caption{\otrace's completeness guarantees organized by attestation type, data controller type, and threat model. In the \hlc[lightblue]{blue} threat model, we assume the provider is trusted/semi-honest and the recipient is covert. In the \hlc[lightred]{red} threat model, we assume the provider and recipient are both covert.}
    \label{tab:completeness}
\end{table*}

\subsection{Protocol}
Figure~\ref{fig:traceability} provides a high-level overview of \otrace. The protocol is defined in terms of concept synchronizations in Appendix~\ref{sec:concepts}. Here, we provide a high-level overview of \otrace's subprotocols:
\begin{description}[leftmargin=0.5cm]
    \item[Data collection setup.] This three-party protocol, involving a consumer, controller (either a provider or recipient), and traceability agent, establishes the consent and traceability terms for the controller's use of the consumer's data.
    \begin{enumerate}
        \item \emph{Consent agreement.} The consumer and controller agree on the terms of data use and traceability.
        \item \emph{Traceability setup.} The consumer introduces the controller to its traceability agent.
        \item \emph{Traceability attestation.} The controller submits a traceability attestation to the traceability agent.
    \end{enumerate}
    \item[Data sharing setup.] This four-party protocol, involving a consumer, provider, recipient, and traceability agent, establishes authorization and terms for data sharing from the provider to the recipient.
    \begin{enumerate}
        \item \emph{Data sharing agreement.} The consumer authorizes data sharing from the provider to the recipient and sets the terms of data sharing.
        \item \emph{Sharing attestation.} The provider and recipient both submit sharing attestations to the traceability agent.
    \end{enumerate}
    \item[Data access.] This three-party protocol, involving a provider, recipient, and traceability agent, traces data-access actions.
    \begin{enumerate}
        \item \emph{Data access.} The recipient accesses data from the provider via an API.
        \item \emph{Access attestation.} The provider and recipient both submit an access attestation to the traceability agent.
    \end{enumerate}
    \item[Data process.] This two-party protocol, involving a controller and traceability agent, traces data-process actions.
    \begin{enumerate}
        \item \emph{Data process.} The controller processes a piece of data.
        \item \emph{Use attestation.} The controller submits a process attestation to the traceability agent.
    \end{enumerate}
    \item[Rights request.] This three-party protocol, involving a consumer, controller, and traceability agent, traces the controller's compliance with consumer rights requests~\cite{drp}.
    \begin{enumerate}
        \item \emph{Rights request.} The consumer submits a rights request to the traceability agent.
        \item \emph{Request attestation.} The traceability agent records a request attestation.
        \item \emph{Forward request.} The traceability agent forwards the request to the controller.
        \item \emph{Request compliance.} The controller complies with the request.
        \item \emph{Rights attestation.} The controller submits a rights attestation to the traceability agent.
    \end{enumerate}
    \item[Traceability access.] This two-party protocol, involving a consumer and traceability agent, allows the consumer to access traceability data.
    \begin{enumerate}
        \item \emph{Traceability access.} The consumer accesses traceability data through the traceability agent.
    \end{enumerate}
\end{description}

\subsection{Threat Model Analysis}

\header{Threat models} Our primary threat model assumes: (1) a trusted traceability agent, (2) trusted or passive data providers, and (3) covert data recipients.\footnote{Whether the data providers are trusted or passive is immaterial for our completeness analysis, since we do not address consumer privacy concerns regarding the provider.} This model captures a range of commercial contexts in which traceability could prove valuable. For example, in the context of open banking, traditional banks (the data providers) already enjoy a reasonable consumer trust as financial stewards. It is, therefore, a natural extension to trust banks to adhere to the protocols they commit to. Moreover, in the United States, for example, banks already operate under stringent regulatory oversight, with agencies like the Consumer Financial Protection Bureau required by statute to take action against deceptive practices if banks stray from their commitments. By contrast, finance management apps (the data recipients) often have not earned the same level of trust from consumers. While they provide valuable services, they may be viewed with more caution due to their newer presence in the market and their frequent reliance on data monetization.

We also consider a second, stronger threat model, which mirrors the primary one but assumes that data providers, like recipients, are covert as well. Indeed, in certain commercial contexts, neither party can be trusted or considered passive. While our traceability protocol is not explicitly tailored to this more challenging model, we will also clarify the guarantees it provides under these conditions.

\header{Types of completeness} The key property we focus on for traceability is completeness, meaning that no attestations are missing, whether intentionally or accidentally. With complete attestations we can achieve full accountability for any privacy violation. In Table~\ref{tab:completeness}, we summarize the completeness guarantees provided by \otrace, organized by attestation type, data controller type, and threat model. The \hlc[lightblue]{blue} threat model is our primary threat model, where we assume the provider is trusted or passive and the recipient is covert. The \hlc[lightred]{red} threat model is the stronger threat model, where both the provider and recipient are assumed to be covert.

Under these threat models, \otrace offers four types of completeness guarantees, depending on the type of attestation:
\begin{itemize}
    \item \emph{assumed} means that completeness is guaranteed because the controller is trusted or passive;
    \item \emph{consumer-dependent} means that completeness is validated by and depends on the consumer;
    \item \emph{covert-secure} means that completeness is provably secure (e.g., guaranteed with high probability);
    \item \emph{covert-accountable} means that completeness is backed by legal accountability.
\end{itemize}

Note that these guarantees are not entirely independent. In fact, covert-accountability can serve as a backstop when other completeness guarantees fall short. For instance, even if assumed completeness is misplaced, the trusted parties often belong to highly regulated industries, so accountability still offers a layer of protection. Similarly, if consumer-dependent completeness is undermined by manipulative dark patterns, consumer protection laws may still provide a remedy.

\header{{\hlc[lightblue]{Trusted/passive provider; covert recipient}}}
In this threat model, we assume providers are either trusted or passive, meaning that completeness for all attestations is trivially guaranteed. Crucially, provider completeness acts as a root of trust for recipient completeness, resulting in a simpler, more scalable protocol.

For recipients, the completeness of consent attestations is consumer-dependent because it hinges on the consumer's ability to ensure these attestations reflect their preferences. No doubt, consumers are notoriously vulnerable to error and manipulation, but some degree of consumer dependence is inevitable since they must provide consent---whether implied or express---for the use of their personal data by controllers. Therefore, a key design goal in practice is to make this process as seamless and intuitive for consumers as possible.


The completeness of sharing, access, and request attestations is covert-secure, achieved using an approach akin to the ``double-entry bookkeeping'' system. In this system, two parties submit attestations, which must be consistent with one another. When one party is trusted or passive, their attestation becomes the root of trust for verifying the other party's attestation. Any inconsistencies between the two can thus be flagged for further review. For sharing and access attestations, the provider acts as the root of trust, while for request attestations, this role is filled by the traceability agent. By contrast, the completeness of process attestations is covert-accountable, since there is no root of trust---interactions occur solely between the recipient and the agent.


\header{\hlc[lightred]{Covert providers and recipients}}
In this stronger threat model, where the provider cannot act as a root of trust, the completeness of the provider's consent attestations becomes consumer-dependent, while all other attestations for both the provider and recipient revert to covert-accountability. Although this shifts the burden of enforcement to the legal system, the technical design can still play a crucial role by enhancing the scalability of legal enforcement.


\header{What the technical design needs from the legal design}
Covert-accountable attestations need teeth to be effective. Without strong enforcement, controllers may lack the incentive to provide them honestly. The primary goal of the legal design is thus to empower enforcers to hold controllers accountable for incompleteness, whether due to unintentional oversight or deliberate misconduct, specifically regarding cover-accountable attestations.

\section{Legal Design for Traceability}
\label{sec:accountability}

This section focuses on the legal dimension of our privacy alignment through data traceability. We start by discussing how existing regulatory frameworks can incentivize or mandate traceability, setting the stage for its adoption. We then propose an accountability framework for policing incompleteness that can be readily implemented under current U.S. privacy laws.

\subsection{Regulatory Frameworks for Adoption}

\header{Voluntary, enforceable codes of conduct}
A key regulatory tool adopted by the FTC to govern online privacy practices is what we has been called a voluntary but enforceable code of conduct. These are promises voluntarily made to consumers, but once made, they become enforceable by the FTC. Such tools have been used over and over again by the FTC in the absence of Congressionally-mandated privacy regulation.
As an alternative to legislation, the FTC called on Internet companies to post privacy policies that would give consumers notice of their privacy rights, hoping that market pressure would lead to fair policies~\cite{ftc1999privacy}. The FTC also made it clear that if a company made a promise it its posted privacy policy, but then failed to keep it, that this breach could be considered a violation of the ``unfair and deceptive practices'' provision of the Federal Trade Commission Act~\cite{ftc_section_5}. 


The FTC can, even without any additional legislative authority, leverage more specific forms of privacy practices. In the case of online advertising, the FTC engaged the advertising industry in a series of negotiations that led to comprehensive codes of conduct governing advertising practices. These codes includes specific data usage and purpose limitation rules, opt-out rights, consumer complaint mechanisms along with industry-funded audits submitted to the FTC on a regular basis~\cite{self-reg-principles}. If the FTC were to push industry sectors to adopt traceability as a new privacy practice, accountability for incompleteness would assume an important role. Similar legally-enforceable codes of conduct are also provided for in the EU General Data Protection Regulation Article 40.

\header{Government regulation} Alternatively, government regulation mandating traceability is a justified response to the market's failure to produce responsible data practices.  Traceability corrects an information asymmetry between controllers and consumers~\cite{acquisti2016economics}. For a competitive market to operate efficiently, consumers need sufficient information to evaluate competitors, including considering their data practices. Otherwise, the market may fail to deliver optimal privacy outcomes. Yet, today's privacy notices fall short and controllers systematically underprotect consumers, whose available legal remedies are often costly, impractical, or nonexistent~\cite{solove2014ftc}. Privacy notices are notoriously difficult and time-consuming to read~\cite{mcdonald2008cost}, while often also providing little meaningful insight into controllers' data practices. Moreover, consumers frequently lack a full understanding of the long-term consequences of sharing personal data~\cite{acquisti2016economics}, leaving them vulnerable to exploitation by controllers who nudge them into accepting terms that are not in their best interest, such as overly broad data-sharing agreements~\cite{acquisti2016economics}.

To address this transparency deficit, we argue that traceability is a natural evolution from the FTC's ``traditional light-touch, ex post information collection approach''~\cite{van2019missing}. Many FTC consent decrees already require regulated parties to maintain records and submit compliance reports to facilitate enforcement~\cite{solove2014ftc}. With traceability, such recording and reporting simply becomes more standardized, systematized, and streamlined. 

Realizing the promise of traceability will depend on a more proactive approach in many sectors, requiring companies to disclose their data practices to consumers upfront. This is similar to how the Securities and Exchange Commission (SEC) requires detailed financial disclosures~\cite{securities_exchange_act_1934} and the Occupational Safety and Health Administration (OSHA) requires comprehensive safety risk disclosures~\cite{osha_act_1970}, regardless of whether an audit or investigation is underway. In fact, the FTC is an outlier in this respect. Most major regulators routinely compel firms to furnish information absent any suspicion of wrongdoing~\cite{van2019missing}.

Traceability can also be seen as an extension of existing data access mandates. Laws like the CCPA and GDPR have highlighted the importance of empowering consumers with data access rights. Increasingly, there is a recognition of the need to extend these rights to consumer agents as well. A notable example is the CFPB's 2023 rule mandating data access within the open banking ecosystem~\cite{doddfrank2010}. Under Section 1033 of the Dodd-Frank Act of 2010, Congress authorized the CFPB to establish rules requiring financial institutions to share consumer data with third parties chosen by the consumer~\cite{cfpb2023}. The 2023 rule builds on this mandate, applying it to the modern context of open banking. This means that if a consumer uses, for instance, a third-party digital assistant that needs access to their financial data, the bank or credit card company must provide API access to that data. Traceability builds upon these access mandates by adding an essential layer of consumer control and accountability to data sharing practices.

\header{Cooperative compliance program}
Beyond self-regulated or government-mandated traceability, a practical step forward in promoting traceability adoption could involve privacy regulators establishing a cooperative compliance program that incentivizes voluntary participation. For example, under OSHA’s Voluntary Protection Programs (VPP), businesses that voluntarily adopt safety and health management systems that exceed OSHA’s baseline requirements are recognized as ``models of excellence'' and are exempt from OSHA's programmed inspections~\cite{osha_vpp}. 
A comparable privacy-focused cooperative compliance program could reduce audit frequency for controllers that voluntarily implement traceability. This creates a network effect: the more participants in the program, the lower the likelihood of audits for each individual participant. Conversely,  controllers who opt out of the program would face more frequent audits, incentivizing broader adoption of traceability.



\subsection{Policing Incompleteness}

\header{Leveraging consumer protection frameworks}
The most straightforward way to police incompleteness may be through existing consumer protection frameworks. These include the FTC's Section 5 authority to prevent ``unfair or deceptive acts or practices''~\cite{ftc_section_5}, along with analogous statutes enforced by other federal agencies, and various state-level ``little FTC Acts''~\cite{butler2011state}. To establish a clear legal hook for holding controllers accountable for incompleteness, those purporting to support traceability should be required to notify consumers that they will fulfill their attestation obligations. Failure to meet these commitments could then serve as grounds for legal action under the aforementioned consumer protection frameworks.

For instance, under the deception prong of the FTC's Section 5 authority, a deceptive trade practice involves: (1) an act of misrepresentation, omission, or practice that (2) misleads a reasonable consumer (3) to the consumer's detriment~\cite{solove2014ftc}. In fact, much of the FTC's privacy jurisprudence rests on a theory of deception through broken promises, typically invoked when a company fails to uphold its own privacy policy~\cite{solove2014ftc}. Incompleteness following a promise of transparency fits squarely within this theory: (1) the controller fails to disclose actual data usage (2) despite promising that consumers will be able to trace this usage, (3) resulting in a breach of the consumer's privacy rights.

The traditional model of privacy enforcement, which relies heavily on investigations and audits, has become increasingly unsustainable in the face of today's vast and complex data ecosystems. Regulators are hard-pressed to police the trillions of data transactions occurring daily among consumers, controllers, and processors. Even when controllers are required to hire third-party auditors to monitor for potential violations, as is often mandated in FTC consent decrees~\cite{solove2014ftc}, the monitoring process remains not only resource-intensive but, worse, often ineffective.


\header{Traceability is key to policing incompleteness at scale} New approaches are clearly needed, and traceability presents a powerful tool for audit-based enforcement to keep pace with the scale and complexity of today's data ecosystems. First, traceability enables crowdsourced detection of potentially wrongful activity. While the FTC actively monitors for unfair and deceptive practices, its capacity is limited by staffing and budget constraints~\cite{solove2014ftc}. As a result the FTC often relies on informal consumer complaints to uncover issues. By correcting the information asymmetry between controllers and consumers, traceability empowers a form of collective monitoring for data misuse. 

Second, traceability reduces evidentiary costs. When implemented proactively, it empowers consumers to provide a pre-established baseline of evidence, alleviating the burden on regulators during investigations, audits, or discovery. This approach benefits controllers as well: by making an upfront investment in traceability, they can potentially avoid the hefty costs of ex-post information collection, or at the very least, streamline any additional information gathering efforts.

Third, traceability simplifies the assignment of liability. Traditionally, privacy disputes have revolved around the simple, dyadic relationship between controller and consumer. However, modern data ecosystems involve an increasingly complex web of intermediaries, including data brokers, ad exchanges, payment processors, and more. This can make it challenging to pinpoint responsibility for privacy violations. By creating a clear, reviewable record of each step in potentially offending data flows, traceability enables precise identification of the parties responsible for violations.


Fourth, traceability helps identify who has been harmed by a privacy violation. One of the major challenges in privacy law is defining the class of individuals affected by a violation~\cite{cofone2023certifying}. Class certification, in particular, is difficult because it requires proving that common issues of law or fact predominate over individual differences. Privacy harms, however, can vary widely across a potential class, and without a clear record of how data has been (mis)used, it becomes challenging to determine who was affected. Traceability addresses this by providing such a record, allowing for precise identification of those impacted.

\section{Discussion}
\label{sec:discussion}



\header{Privacy-alignment for third-party data sharing}
Viewed through the lens of our new framework for threat modeling and privacy analysis, \otrace makes for a privacy-aligned environment for third-party data sharing. Our approach first introduces a different concept of privacy than what is traditionally found in the technical literature. While conventional notions of privacy emphasize keeping data hidden, our approach focuses on transparency and control over data sharing. \otrace's primary goal is thus to empower consumers with the knowledge of who has their personal data, what it is being used for, and whom it is being shared with---as provided through attestations.

In this context, privacy means that these attestations are complete---no attestations are missing, whether intentionally or inadvertently. Our approach also shifts from traditional security analysis frameworks in two key ways: First, it reflects the practical reality that parties in commercial settings vary in trustworthiness. Second, it recognizes that privacy guarantees can be achieved not only through technical mechanisms, but also through legal enforcement designed specifically to dovetail with technical design, addressing identify risks in a clear threat model.

Applying this framework to the design of \otrace clarifies the contours of where technical guarantees end and where legal guarantees begin. We begin with a practical assumption common in many data-sharing contexts: that data providers can generally be trusted. \otrace leverages this trust by positioning providers as roots of trust, implementing technical checks on the actions of less-trusted, covert data recipients who might attempt to evade detection. In particular, these checks extend across attestations for data-sharing agreements between providers and recipients, recipients' data accesses, and their compliance with data rights requests. \otrace's design critically provides technical completeness guarantees that can be enforced perfectly and automatically, without the need for complex, resource-intensive technical infrastructure that would otherwise be required in the absence of trusted providers.

Our analysis also reveals a key gap: ensuring the completeness of attestations regarding data recipients' local processing. Since these attestations occur solely  between the recipient and the traceability agent---without a root of trust---rather than falling back to heavyweight, technical solutions, we turn to covert-accountability as a security guarantee. Our legal design then shows how the completeness of local processing attestations can be enforced under existing consumer protection frameworks. Furthermore, we argue how \otrace's technical affordances can provide more confident, scalable regulatory oversight.

\header{Making consumer control meaningful}
Another key set of challenges relates to how consumers interact with traceability agents. We offer several  design considerations, which are the focus of ongoing studies. To start, privacy consent is a cornerstone to many privacy frameworks, including ours. Yet, we know all too well that it is fraught with challenges: transaction costs are prohibitively high~\cite{mcdonald2008cost}, controllers employ ``dark patterns'' to ``coerce, wheedle, and manipulate people to grant [consent,]''~\cite{richards2018pathologies}, and the list goes on. Nevertheless, we argue that our traceability protocol can help shift privacy consent from legal fiction to meaningful practice by promoting granularity and consistency across consent experiences. This means that consumers not only have the freedom to specify which data types and purposes they consent to, but also that consent experiences feel familiar even across varied controllers, thereby reducing the cognitive load of granularity.


Additionally, there is the question of how traceability data will be presented to consumers. Simply providing a raw dump of attestations is unlikely to empower consumers. Instead, traceability access should be tailored to individual preferences and needs. In some cases, a ``push'' model may be more appropriate, where agents proactively send notifications to consumers, such as when a violation occurs. In others, a ``pull'' model might be preferable, allowing consumers to access and visualize specific data through a traceability dashboard.

\header{Trusting the agents}
\otrace also rests on the critical assumption that traceability agents are trustworthy. However, this assumption is not without its challenges. Legitimate concerns arise, for instance, when agents are run by for-profit companies. These companies, by their very nature, prioritize profit, which can lead to decisions that favor their financial interests over those of consumers. 

This begs the question of how to hold agents themselves accountable. One approach is for the government to enlist private entities as agents or to transform them into ``mini'' consumer protection regulators~\cite{van106new}. Regulators might look to existing consumer protection advocates who can build tools and cooperative analytic platforms to help individual consumers aggregate their own traceability data into collective insights on the behavior of various firms in the marketplace (see, e.g., Consumer Report's Permission Slip project~\cite{permission-slip}).

Another approach to strengthening agent accountability is to impose a fiduciary duty on them~\cite{balkin2015information}. This legal obligation would require agents to prioritize consumers' interests over their own financial gain. Fiduciary duties have traditionally been applied to professionals in positions of trust and power, such as doctors and lawyers, to prevent exploitation and ensure that trust is not abused. By holding traceability agents to this heightened standard of care, any breach of their fiduciary duty would expose them to liability, thereby reinforcing their accountability.
\section{Conclusion}
\label{sec:conclusion}

Traceability revives the core privacy principle of transparency, first articulated in 1973, which advocates that consumers should have visibility into what data is being collected about them and how it is being used~\cite{us_dhhs_1973}. By re-implementing this principle in a modern technical context, and showing how its technical guarantees can be combined with legal measures, we demonstrate a means of designing accountable systems for large-scale third-party data sharing ecosystems systems that meet a formally defined privacy threat model. We hope that this \emph{legal-technical alignment} approach based on a principled threat model analysis can provide increased confidence in data governance systems and ultimately restore consumer trust in modern data environments. 

\section*{Acknowledgements}
This is the work of the entire OTrace team at MIT including Daniel Jackson, project co-Principal Investigator, Ashar Farooq, Nicola Tatyana Lawford, Eagon Meng, and Jian-Ming Chen. We extend special thanks to alums Quinn Magendanz and Dean Wen for their contributions to \otrace. We also thank Hari Balakrishnan, Sukhi Gulati-Gilbert, Ilaria Liccardi, Aileen Nielsen, Erica Radler, and Lily Tsai for their prior research on accountable systems and feedback on earlier drafts. This work was funded in part by the NSF CCF-2131541 (DASS: Legally Accountable Cryptographic Computing Systems (LAChS)) and the MIT Future of Data Initiative. Work by Kevin Liao was supported by fellowships from the 28twelve Foundation and the Berkman Klein Center for Internet \& Society.

\bibliographystyle{ACM-Reference-Format.bst}
\bibliography{bibs/references.bib}

\appendix
\appendix

\section{Traceability Concepts}
\label{sec:concepts}

\onecolumn

\begin{itemize}[leftmargin=*, noitemsep]
    \item[] \textbf{concept} introduction [Controller, Subject, TraceService]
    \item[] \textbf{purpose} to introduce a controller to the subject’s traceability service
    \item[] \textbf{state}
    \begin{itemize}[noitemsep]
        \item[] \textcolor{myforestgreen}{// fields}
        \item[] subject: Introduction $\rightarrow$ Subject
        \item[] controller: Introduction $\rightarrow$  Controller
        \item[] traceService: Introduction $\rightarrow$ TraceService
    \end{itemize}
    \item[] \textbf{actions}
    \begin{itemize}[noitemsep]
        \item[] \textcolor{myforestgreen}{// called by subject}
        \item[] introduce(s: Subject, c: Controller, t: TraceService) $\rightarrow$ i: Introduction
        \begin{itemize}[noitemsep]
            \item[] creates fresh introduction i such that
            \begin{itemize}[noitemsep]
                \item[] i.subject = s
                \item[] i.controller = c
                \item[]	i.traceService = t
            \end{itemize}
        \end{itemize}
    \end{itemize}
\end{itemize}
\noindent\rule{\textwidth}{1pt}
\begin{itemize}[leftmargin=*, noitemsep]
    \item[] \textbf{concept} consent [Controller, Data, Purpose, Subject]
    \item[] \textbf{purpose} to form and manage consent agreements governing a controller’s use of subject’s personal data
    \item[] \textbf{state}
    \begin{itemize}[noitemsep]
        \item[] \textcolor{myforestgreen}{// type definitions}
        \item[] Term: (Data, Purpose)
        \item[] ConsentStatus: REQUESTED | ACCEPTED | DENIED | REVOKED | EXPIRED
        \item[] controller: Consent $\rightarrow$ Controller
        \item[] subject: Consent $\rightarrow$ Subject
        \item[] terms: Consent $\rightarrow$ set Term
        \item[] expiry: Consent $\rightarrow$ Timestamp
        \item[] status: Consent $\rightarrow$ ConsentStatus
    \end{itemize}
    \item[] \textbf{actions}
    \begin{itemize}[noitemsep]
        \item[] \textcolor{myforestgreen}{// called by controller}
        \item[] request(c: Controller, s: Subject, ts: set Term, e: Timestamp) $\rightarrow$ consent: Consent
        \begin{itemize}[noitemsep]
            \item[] creates fresh consent such that
            \begin{itemize}[noitemsep]
                \item[] consent.controller = c
                \item[] consent.subject = s
                \item[] consent.terms = ts
                \item[] consent.expiry = e
                \item[] consent.status = REQUESTED
            \end{itemize}
        \end{itemize}
    \end{itemize}
    \begin{itemize}[noitemsep]
        \item[] \textcolor{myforestgreen}{// called by subject}
        \item[] accept(s: Subject, c: Consent) $\rightarrow$ c: Consent
        \begin{itemize}[noitemsep]
            \item[] requires c.status = REQUESTED
            \item[] c.status = ACCEPTED
        \end{itemize}
    \end{itemize}
    \begin{itemize}[noitemsep]
        \item[] deny(s: Subject, c: Consent) $\rightarrow$ c: Consent
        \begin{itemize}[noitemsep]
            \item[] requires c.status = REQUESTED and c.subject = s
            \item[] c.status = DENIED
        \end{itemize}
    \end{itemize}
    \begin{itemize}[noitemsep]
        \item[] revoke(s: Subject, c: Consent) $\rightarrow$ c: Consent
        \begin{itemize}[noitemsep]
            \item[] requires c.status = ACCEPTED and c.subject = s
            \item[] c.status = REVOKED
        \end{itemize}
    \end{itemize}    
    \begin{itemize}[noitemsep]
        \item[] \textcolor{myforestgreen}{// background action}
        \item[] expire(c: Consent)
        \begin{itemize}[noitemsep]
            \item[] requires c.status = ACCEPTED and c.expiry is before now
            \item[] c.status = EXPIRED
        \end{itemize}
    \end{itemize}
    \begin{itemize}[noitemsep]
        \item[] \textcolor{myforestgreen}{// called by controller}
        \item[] permit(c: Controller, s: Subject, t: Term)
        \begin{itemize}[noitemsep]
            \item[] requires some consent c where c.status = ACCEPTED and t in c.terms
        \end{itemize}
    \end{itemize}    
    \item[] \textbf{operational principle}
        \begin{itemize}[noitemsep]
        \item[] \textcolor{myforestgreen}{// after a controller requests consent for use of a subject’s data under some terms, and the subject accepts that consent, then so long as the subject does not revoke the consent and the consent does not expire, the controller is permitted to use the subject’s data on those terms}
        \item[] after
        \begin{itemize}[noitemsep]
            \item[] x = request(c, s, ts, e)
            \item[] accept(s, x)
        \end{itemize}
        \item[] then unless
        \begin{itemize}[noitemsep]
            \item[] expire(x) or revoke(s, x)
        \end{itemize}
        \item[]  then for any t in ts can do
        \begin{itemize}[noitemsep]
            \item[] permit(c, s, t)
        \end{itemize}
    \end{itemize} 
\end{itemize}
\noindent\rule{\textwidth}{1pt}
\begin{itemize}[leftmargin=*, noitemsep]
    \item[] \textbf{concept} authorization [Data, Provider, Recipient, Subject]
    \item[] \textbf{purpose} to authorize a provider to share data about a subject with a recipient
    \item[] \textbf{state}
    \begin{itemize}[noitemsep]
        \item[] subject: Authorization $\rightarrow$ Subject
        \item[] provider: Authorization $\rightarrow$ Provider
        \item[] recipient: Authorization $\rightarrow$ Recipient
        \item[] data: Authorization $\rightarrow$ set Data
        \item[] expiration: Authorization $\rightarrow$ Timestamp
    \end{itemize}
    \item[] \textbf{actions}
    \begin{itemize}[noitemsep]
        \item[] \textcolor{myforestgreen}{// called by subject}
        \item[] authorize(s: Subject, p: Provider, r: Recipient, d: set Data, e: Expiration) $\rightarrow$ a: Authorization
        \begin{itemize}[noitemsep]
            \item[] creates fresh authorization a such that
            \begin{itemize}[noitemsep]
                \item[] a.subject = s
                \item[] a.provider = p
                \item[] a.recipient = r
                \item[] a.data = d
                \item[] a.expiration = e
            \end{itemize}
        \end{itemize}
        \item[] revoke(s: Subject, a: Authorization) $\rightarrow$ a: Authorization
        \begin{itemize}[noitemsep]
            \item[] a.expiration = Timestamp.now()
        \end{itemize}
    \end{itemize}
\end{itemize}
\noindent\rule{\textwidth}{1pt}
\begin{itemize}[leftmargin=*, noitemsep]
    \item[] \textbf{concept} dataUse [Controller, Data, Basis, Subject]
    \item[] \textbf{purpose} to use data about a subject for a specific purpose in association with a [legal] basis
    \item[] \textbf{state}
    \begin{itemize}[noitemsep]
        \item[] \textcolor{myforestgreen}{// fields}
        \item[] controller: DataUse $\rightarrow$ Controller
        \item[] subject: DataUse $\rightarrow$ Subject
        \item[] data: DataUse $\rightarrow$ Data
        \item[] operation: DataUse $\rightarrow$ Operation
        \item[] basis: DataUse $\rightarrow$ Basis
    \end{itemize}
    \item[] \textbf{actions}
    \begin{itemize}[noitemsep]
        \item[] \textcolor{myforestgreen}{// called by controller}
        \item[] use(c: Controller, s: Subject, d: Data, o: Operation, b: Basis) $\rightarrow$ u: DataUse
        \begin{itemize}[noitemsep]
            \item[] creates fresh data use u such that
            \begin{itemize}[noitemsep]
                \item[] u.controller = c
                \item[] u.subject = s
                \item[] u.data = d
                \item[] u.operation = o
                \item[] u.basis = b
            \end{itemize}
        \end{itemize}
        \item[] \textcolor{myforestgreen}{// called by subject}
        \item[] getBasis(u: DataUse) $\rightarrow$ u.basis: Basis
    \end{itemize}
    \item[] \textbf{operational principle}
    \begin{itemize}[noitemsep]
        \item[] Pair use and basis
    \end{itemize}
\end{itemize}
\noindent\rule{\textwidth}{1pt}
\begin{itemize}[leftmargin=*, noitemsep]
    \item[] \textbf{concept} dataSubjectRequest [Controller, Subject]
    \item[] \textbf{purpose} to make a data subject request (e.g., access, correction, deletion, or opting out of sales) to a controller
    \item[] \textbf{state}
    \begin{itemize}[noitemsep]
        \item[] \textcolor{myforestgreen}{// type definitions}
        \item[] RequestType: ACCESS | CORRECT | DELETE | OPTOUT | \dots
        \item[] RequestStatus: SENT | RECEIVED | COMPLETED | DENIED
        \item[] \textcolor{myforestgreen}{// fields}
        \item[] subject: DSR $\rightarrow$ Subject
        \item[] controller: DSR $\rightarrow$ Controller
        \item[] type: DSR $\rightarrow$ RequestType
        \item[] status: DSR $\rightarrow$ RequestStatus
    \end{itemize}
    \item[] \textbf{actions}
    \begin{itemize}[noitemsep]
        \item[] \textcolor{myforestgreen}{// called by subject}
        \item[] sendRequest(s: Subject, c: Controller, t: RequestType) $\rightarrow$ r: DSR
        \begin{itemize}[noitemsep]
            \item[] creates fresh data subject request r such that
            \begin{itemize}[noitemsep]
                \item[] r.subject = s
                \item[] r.controller = c
                \item[] r.type = t
                \item[] r.status = SENT
            \end{itemize}
        \end{itemize}
        \item[] \textcolor{myforestgreen}{// called by controller}
        \item[] receiveRequest(c: Controller, r: DSR) $\rightarrow$ r: DSR
        \begin{itemize}[noitemsep]
            \item[] requires r.status = SENT
            \item[] r.status = RECEIVED
        \end{itemize}
        \item[] completeRequest(c: Controller, r: DSR) $\rightarrow$ r: DSR
        \begin{itemize}[noitemsep]
            \item[] requires r.status = RECEIVED
            \item[] r.status = COMPLETED
        \end{itemize}
        \item[] denyRequest(c: Controller, r: DSR) $\rightarrow$ r: DSR
        \begin{itemize}[noitemsep]
            \item[] requires r.status = RECEIVED
            \item[] r.status = DENIED
        \end{itemize}
    \end{itemize}
\end{itemize}
\noindent\rule{\textwidth}{1pt}
\begin{itemize}[leftmargin=*, noitemsep]
    \item[] \textbf{concept} attestation [Action, Party]
    \item[] \textbf{purpose} to allow parties to attest to report actions performed at a particular time
    \item[] \textbf{state}
    \begin{itemize}[noitemsep]
        \item[] \textcolor{myforestgreen}{// fields}
        \item[] party: Attestation $\rightarrow$ Party
        \item[] action: Attestation $\rightarrow$ Action
        \item[] timestamp: Attestation $\rightarrow$ Timestamp
        \item[] \textcolor{myforestgreen}{// global state}
        \item[] attestations: set Attestation
    \end{itemize}
    \item[] \textbf{actions}
    \begin{itemize}[noitemsep]
        \item[] \textcolor{myforestgreen}{// called by party}
        \item[] attest(p: Party, a: Action)
        \begin{itemize}[noitemsep]
            \item[] creates fresh attestation att such that
            \begin{itemize}[noitemsep]
                \item[] att.party = p
                \item[] att.action = a
                \item[] att.timestamp = Timestamp.now()
                \item[] attestations.add(att)
            \end{itemize}
        \end{itemize}
        \item[] \textcolor{myforestgreen}{// called by subject}
        \item[] getAll() $\rightarrow$ attestations: set Attestation
        \item[] \textcolor{myforestgreen}{// called by controller}
        \item[] upToDate(startDate: Date, endDate: Date, as: set Attestation)
        \begin{itemize}[noitemsep]
            \item[] generate an object attesting that as includes all actions between startDate and endDate
        \end{itemize}
    \end{itemize}
    \item[] \textbf{operational principle}
    \begin{itemize}[noitemsep]
        \item[] after attest(p, a) $\rightarrow$ att, then att in getAll()
    \end{itemize}
\end{itemize}
\noindent\rule{\textwidth}{1pt}

\begin{itemize}[leftmargin=*, noitemsep]
    \item[] \textcolor{myforestgreen}{// attestation setup synchronizations}
    \item[] \textbf{when}
    \begin{itemize}[noitemsep]
        \item[] introduction.introduce(s: Subject, c: Controller, t: TraceService) $\rightarrow$ i: Introduction
    \end{itemize}
    \item[] \textbf{sync}
    \begin{itemize}[noitemsep]
        \item[] attestation.attest(s, i)
        \item[] attestation.attest(c, i)
    \end{itemize}
\end{itemize}

\noindent\rule{\textwidth}{1pt}

\begin{itemize}[leftmargin=*, noitemsep]
    \item[] \textcolor{myforestgreen}{// consent attestation synchronizations}
    \item[] \textbf{when}
    \begin{itemize}[noitemsep]
        \item[] consent.offer(c: Controller, s: Subject, t: set Term, e: Timestamp) $\rightarrow$ consent: Consent
    \end{itemize}
    \item[] \textbf{sync}
    \begin{itemize}[noitemsep]
        \item[] attestation.attest(c, consent)
    \end{itemize}
    
    \item[] \textbf{when}
    \begin{itemize}[noitemsep]
        \item[] consent.accept(s: Subject, c: Consent) $\rightarrow$ c': Consent
    \end{itemize}
    \item[] \textbf{sync}
    \begin{itemize}[noitemsep]
        \item[] attestation.attest(s, c')
        \item[] attestation.attest(c'.controller, c')
    \end{itemize}
    
    \item[] \textbf{when}
    \begin{itemize}[noitemsep]
        \item[] consent.deny(s: Subject, c: Consent) $\rightarrow$ c': Consent
    \end{itemize}
    \item[] \textbf{sync}
    \begin{itemize}[noitemsep]
        \item[] attestation.attest(s, c')
        \item[] attestation.attest(c'.controller, c')
    \end{itemize}

    \item[] \textbf{when}
    \begin{itemize}[noitemsep]
        \item[] consent.revoke(s: Subject, c: Consent) $\rightarrow$ c': Consent
    \end{itemize}
    \item[] \textbf{sync}
    \begin{itemize}[noitemsep]
        \item[] attestation.attest(s, c')
        \item[] attestation.attest(c'.controller, c')
    \end{itemize}
\end{itemize}

\noindent\rule{\textwidth}{1pt}

\begin{itemize}[leftmargin=*, noitemsep]
    \item[] \textcolor{myforestgreen}{// authorization attestation synchronizations}
    \item[] \textbf{when}
    \begin{itemize}[noitemsep]
        \item[] authorization.authorize(s: Subject, p: Provider, r: Recipient, d: set Data, e: Expiration) $\rightarrow$ a: Authorization
    \end{itemize}
    \item[] \textbf{sync}
    \begin{itemize}[noitemsep]
        \item[] attestation.attest(s, a)
        \item[] attestation.attest(p, a)
        \item[] attestation.attest(r, a)
    \end{itemize}

    \item[] \textbf{when}
    \begin{itemize}[noitemsep]
        \item[] authorization.revoke(s: Subject, a: Authorization) $\rightarrow$ a': Authorization
    \end{itemize}
    \item[] \textbf{sync}
    \begin{itemize}[noitemsep]
        \item[] attestation.attest(s, a')
        \item[] attestation.attest(a'.provider, a')
        \item[] attestation.attest(a'.recipient, a')
    \end{itemize}
\end{itemize}

\noindent\rule{\textwidth}{1pt}

\begin{itemize}[leftmargin=*, noitemsep]
    \item[] \textcolor{myforestgreen}{// data use attestation synchronizations}
    \item[] \textcolor{myforestgreen}{// attestation upToDate synchronization: applicable when investigating an action based on a consent, not a law}
    \item[] \textbf{when}
    \begin{itemize}[noitemsep]
        \item[] dataUse.use(c: Controller, s: Subject, d: Data, b: Basis) $\rightarrow$ u: DataUse
    \end{itemize}
    \item[] \textbf{sync}
    \begin{itemize}[noitemsep]
        \item[] attestation.attest(c, u)
        \item[] T = Time.now()
        \item[] attestation.upToDate(d1: Basis.timestamp, d2: T, s: set of attestations that are considered up to date)
    \end{itemize}
\end{itemize}

\noindent\rule{\textwidth}{1pt}

\begin{itemize}[leftmargin=*, noitemsep]
    \item[] \textcolor{myforestgreen}{// data subject request attestation synchronizations}
    
    \item[] \textbf{when}
    \begin{itemize}[noitemsep]
        \item[] dataSubjectRequest.sendRequest(s: Subject, c: Controller, t: RequestType) $\rightarrow$ r: DSR
    \end{itemize}
    \item[] \textbf{sync}
    \begin{itemize}[noitemsep]
        \item[] attestation.attest(s, r)
    \end{itemize}

    \item[] \textbf{when}
    \begin{itemize}[noitemsep]
        \item[] dataSubjectRequest.receiveRequest(c: Controller, r: DSR) $\rightarrow$ r': DSR
    \end{itemize}
    \item[] \textbf{sync}
    \begin{itemize}[noitemsep]
        \item[] attestation.attest(c, r')
    \end{itemize}

    \item[] \textbf{when}
    \begin{itemize}[noitemsep]
        \item[] dataSubjectRequest.completeRequest(c: Controller, r: DSR) $\rightarrow$ r': DSR
    \end{itemize}
    \item[] \textbf{sync}
    \begin{itemize}[noitemsep]
        \item[] attestation.attest(c, r')
    \end{itemize}

    \item[] \textbf{when}
    \begin{itemize}[noitemsep]
        \item[] dataSubjectRequest.denyRequest(c: Controller, r: DSR) $\rightarrow$ r': DSR
    \end{itemize}
    \item[] \textbf{sync}
    \begin{itemize}[noitemsep]
        \item[] attestation.attest(c, r')
    \end{itemize}
\end{itemize}

\noindent\rule{\textwidth}{1pt}

\twocolumn

\end{document}